\documentclass[twocolumn,showpacs,aps,prb]{revtex4}
\usepackage{graphics}

\begin{document}
\title{Anomalous Spectral weight in photoemission spectra of the hole
doped Haldane chain $Y_{2-x}Sr_xBaNiO_5$.}

\author{Y. Fagot-Revurat and D. Malterre\\}
\affiliation{Laboratoire de Physique des Mat\'eriaux,
Universit\'e Henri Poincar\'e, Nancy I - B.P. 239\\
F-54506 Vand\oe uvre-l\`es-Nancy, France}

\author{F.-X. Lannuzel, E. Janod and C. Payen\\}
\affiliation{Institut des Mat\'eriaux Jean Rouxel,Universit\'e de Nantes-CNRS,\\
F-44322 Nantes Cedex 3, France}

\author{L. Gavioli and F. Bertran}
\affiliation{Laboratoire pour l'Utilisation du Rayonnement Electromagn\'etique,Universit\'e Paris-Sud\\
F-91405 Orsay Cedex, France\\}

\date{\today}
%
%
\begin{abstract}

In this paper, we present photoemission experiments on the hole
doped Haldane chain compound $Y_{2-x}Sr_xBaNiO_5$. By using the
photon energy dependence of the photoemission cross section, we
identified the symmetry of the first ionisation states (d type).
Hole doping in this system leads to a significant increase in the
spectral weight at the top of the valence band without any change
in the vicinity of the Fermi energy. This behavior, not observed
in other charge transfer oxides at low doping level, could result
from the inhomogeneous character of the doped system and from a Ni
3d-O 2p hybridization enhancement due to the shortening of the
relevant Ni-O distance in the localized hole-doped regions.
\end{abstract}

\maketitle
\preprint{submitted
to Phys. Rev. B}

\section{introduction}

Oxides of 3d transition metals have attracted considerable
attention in the last two decades because they exhibit very varied
and fascinating properties (metal-insulator transition, high Tc
superconductivity, intricating magnetic properties). These
behaviors partly result from an interplay of low dimensionality
and electronic correlations. Many transition metal oxides exhibit
a two dimensional electronic structure. Due to its peculiar
crystal structure, $Y_2BaNiO_5$ is essentially a one dimensional
(1D) divalent nickel oxide. It crystallizes in a body-centered
orthorhombic structure within the Immm space group\cite{Amador}.
The structure is characterized by linear chains of NiO$_6$
octahedra sharing corners along the z direction. The octahedra are
compressed along the chains resulting in two short Ni-O distances
(0.188 nm) and four longer Ni-O distances (0.219 nm)\cite{Amador}.
This unusual contraction and vanishing interactions between
neighboring chains lead to a quasi-one-dimensional electronic
structure\cite{mattheiss}. The hybridization between Ni-3d and
O-2p states is strong along the chain direction. Yet, $Y_2BaNiO_5$
has an insulating ground state due to exchange and correlation
effects. The superexchange antiferromagnetic coupling between
neighboring S=1 spins carried by Ni$^{2+}$ ions in $Y_2BaNiO_5$
yields a quantum spin liquid ground state\cite{dar}. Large quantum
spin fluctuations associated with the 1D character of the Ni-O-Ni
network prevent the formation of an ordered magnetic state. The
spin liquid can be described as macroscopically coherent quantum
state\cite{affl} with a Haldane gap in the spin excitation
spectrum \cite{dar,dit}. Each S=1 spin can be considered as the
triplet state formed from the two ferromagnetically coupled S=1/2
holes of Ni$^{2+}$. Two S=1/2 holes of neighboring sites form a
singlet state. The ground state is then described in the
well-known valence bond solid (VBS) picture. Excitations are
triplet states obtained by breaking valence bonds. Hole doping in
this spin liquid state results in interesting behaviors.
Low-energy spin excitations are characterized by incommensurate
spin density modulation around impurities, and the dynamical spin
structure factor reveals spectral weight in the Haldane gap
\cite{dit,dag,xu}.

>From an electronic point of view, $Y_2BaNiO_5$, like most of the
Ni and Cu oxides, belongs to the charge transfer regime in the
Zaanen-Sawatzky-Allen diagram \cite{zaa}. Then if the low-energy
properties usually associated with spin excitations could be very
different, the charge excitations and in particular the effect of
doping should look very similar in these families. The
photoemission spectrum of homogeneously hole doped systems only
exhibits small spectral weight modifications with doping whereas
large spectral weight transfer is observed in the inverse
photoemission spectrum. In charge transfer oxides, a p symmetry is
expected for the first ionization states. Thus, the substitution
of Sr or Ca divalent ions for trivalent Y in $Y_{2-x}Sr_xBaNiO_5$
should yield holes in the ligand p band. Nevertheless,
hybridization between Ni-3d states and O-2p states can modify this
picture by shifting highly hybridized $d^8\underline{L}$ states
toward the top of the O-2p band ($\underline{L}$ represents a hole
on the ligand). As a consequence, the first ionization states
probed by photoemission should have a 3d character. This
spectroscopy, by measuring the electron-removal spectrum, can give
information about the symmetry of the first ionization states and
about the modification of electronic properties in hole doped
compounds. The electronic structure of hole doped
 $Y_2BaNiO_5$ has been investigated with several methods including
 x-ray absorption spectroscopy (XAS)\cite{dit}, optical
conductivity\cite{ito}, photoemission and inverse
photoemission\cite{mai2}. These experiments show that doped holes
have either $O(2p_z)$ or mixed $Ni3d_{3z^2-r^2}-O(2p_z)$
character.  Hole doping introduces new empty localized states in
the charge gap region without any spectral weight very close to
the Fermi level. The doped carriers are trapped in local bound
states and should therefore locally distort the electronic
structure only in the vicinity of their location giving rise to a
possible non-homogeneous doping at the nanoscopic scale
\cite{xu,ito}. Therefore, the system does not become metallic upon
doping,and no charge ordering is observed.

In this paper, we investigate the electronic structure of
$Y_{2-x}Sr_xBaNiO_5$ for x=0, 0.1 and 0.2 by photoemission
spectroscopy. By using the photon energy dependence of the
spectral density, we confirm that the first electron removal
excitations have mainly a Ni-3d character. Moreover, these
measurements illustrate the effect of hole doping in this system
by evidencing an increase in the spectral weight of d symmetry at
the top of the valence band.  In contrast to most of hole doped
cuprates and nickelates which only present significant spectral
modifications for the unoccupied electronic states, a large
spectral weight modification is observed in the photoemission
spectra of $Y_{2-x}Sr_xBaNiO_5$ for doping level as low as x=0.1.
This singular behavior probably results from the large $Ni3d-O2p$
hybridization along the chain direction and from the inhomogeneous
character of the doped system.

%
%
\section{experimental details}
\label{experimental details}

Polycrystalline samples of $Y_{2-x}Sr_xBaNiO_5$ (x=0.0, 0.10 and
0.20) were prepared by standard solid state reactions and
characterized by X-ray powder diffraction, microprobe and
thermogravimetric analysis. Samples are single-phased,
homogeneously doped and stoichiometric in oxygen\cite{jalcom}.
Photoemission experiments have been carried out on the SU3
beamline of the French synchrotron facilities (LURE). The photon
energy used was in the range extending from 60 eV to 200 eV. The
sample surfaces were scraped in situ with a diamond file to obtain
clean surfaces under the ultrahigh vacuum of low $10^{-10}$ Torr.
The cleanliness of the surfaces was checked by the lack of
contaminant in the valence band spectrum and by X-ray
photoemission core level spectra. The Fermi energy was calibrated
by measuring the metallic edge of a Cu foil close to the sample.
All measurements were carried out at room temperature with an
energy resolution better than 50 meV.
%
%
\section{Results and discussion}

\begin{figure}
\begin{center}
\scalebox{0.4}{\includegraphics{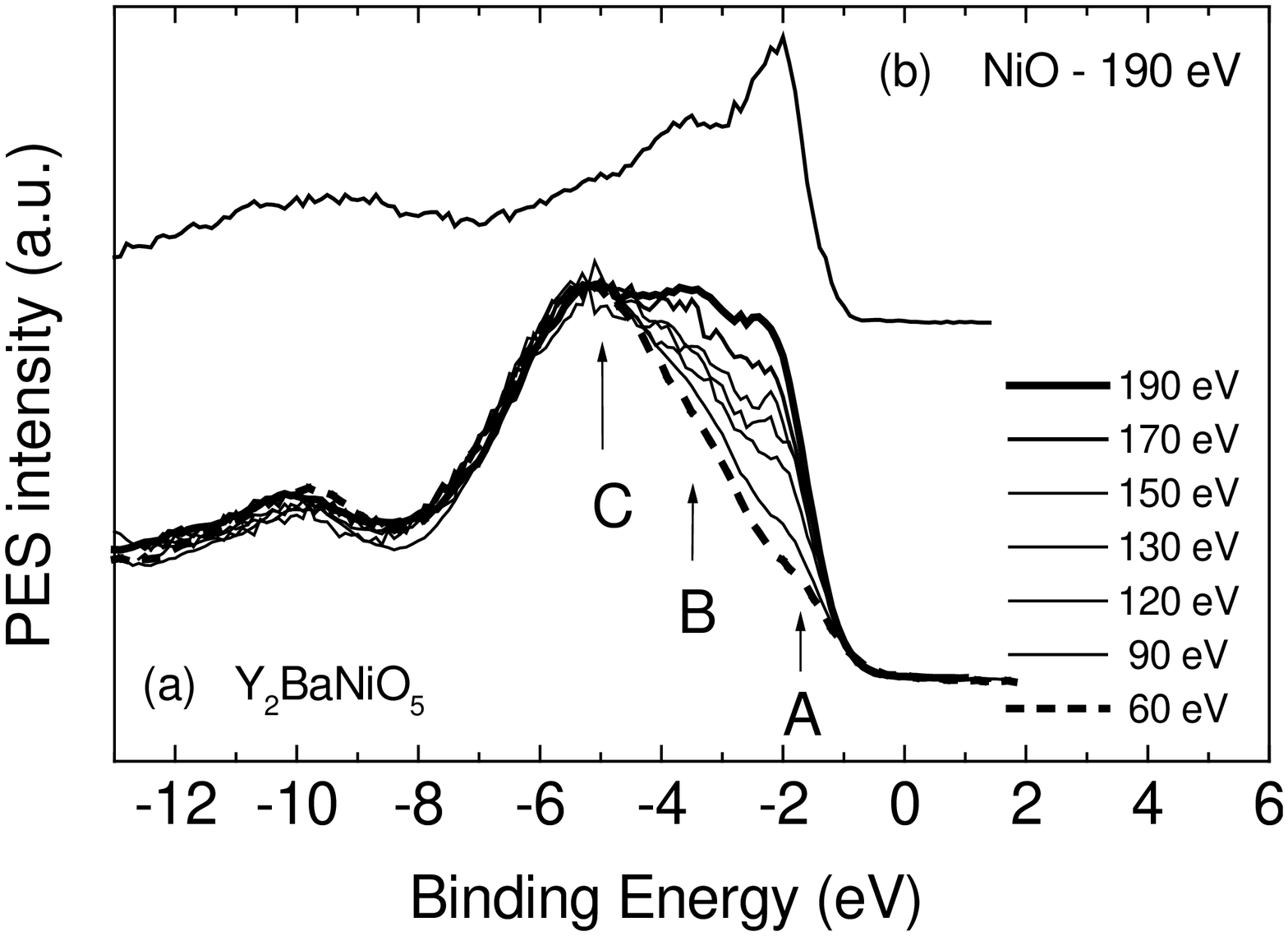}}
\end{center}
\caption{\label{one}(a)Angle-integrated photoemission spectra of
$Y_2BaNiO_5$ for several photon energies in the 60 eV-200 eV
range. The spectra are arbitrary normalized on the maximum of the
C structure.(b) h$\nu$=190 eV- spectrum  of NiO.}
\end{figure}
In figure \ref{one} we present angle-integrated photoemission
spectra of $Y_2BaNiO_5$ for photon energy varying between 60 eV
and 200 eV. These spectra exhibit three features at $E\approx$
-2.0 eV, $E\approx$ -3.5 eV and $E\approx$ -5.0 eV, hereafter
referred to as A, B and C, respectively. The peak at -10 eV,
usually observed in Ni oxides could originate from multielectronic
effects and/or impurity features \cite{mai1}. The spectra are
arbitrarily normalized on the C structure. With this
normalization, a very strong photon energy dependence of the A
feature intensity is clearly observed. Owing to the different
photon energy dependence of d and p states cross section, this
behavior suggests that the A feature is dominated by the Ni-3d
states whereas the B and C feature have likely p character. This
result is in complete agreement with the previous measurements
carried out in the Ca substituted oxide ($Y_{2-x}Ca_xBaNiO_5$)
with HeI ($h\nu$=21.2 eV) and Mg-$K_{\alpha}$($h\nu$=1253.6 eV)
excitations \cite{mai1} which exhibit the same trend. Moreover,
spin polarized band calculations in the local density
approximation confirm this statement by showing that the C feature
can be associated with O2p-Ni3d bonding states with dominant
oxygen character whereas the states in the energy region of the B
feature correspond to oxygen 2p-2p interactions with a non-bonding
character with respect to Ni3d-O2p interactions\cite{mai1,nov}. A
satisfactory agreement between band calculation and photoemission
spectra are found for these two kinds of states \cite{mai1,nov}.
The top of the valence band in the calculations corresponds to
anti-bonding states with a dominant Ni-3d contribution. Recent
LSDA+U calculations, which predict a quite correct energy gap
value, point out the highly hybridized character of these
states\cite{nov}. A significant disagreement between band
calculated density of states and photoemission spectrum are
exhibited for these states. This is not surprising since the A
structure represents excitations of correlated d states which are
poorly described by one-electron (ground state) approaches. Such a
valence spectrum is usually observed in divalent Ni oxides like
NiO\cite{fuji}, as observed in Fig.\ref{one}(b) where the
photoemission spectrum of NiO recorded with 190 eV photon-energy
is presented. In NiO, the main structure close to -2eV has been
interpreted to be due to $d^8\underline{L}$ states. The Ni-3d
derived structures are less resolved for $Y_2BaNiO_5$ than for NiO
(at the same photon energy) because the transition metal to oxygen
atomic ratio is five time larger in the $Y_2BaNiO_5$ compound.
However, an increase in the absolute intensity of this feature can
be evidenced by lowering the dimensionality in NiO (3D),
$La_2NiO_4$ (2D) and $Y_2BaNiO_5$ (1D)\cite{mai2}.

\begin{figure}
\begin{center}
\scalebox{0.4}{\includegraphics{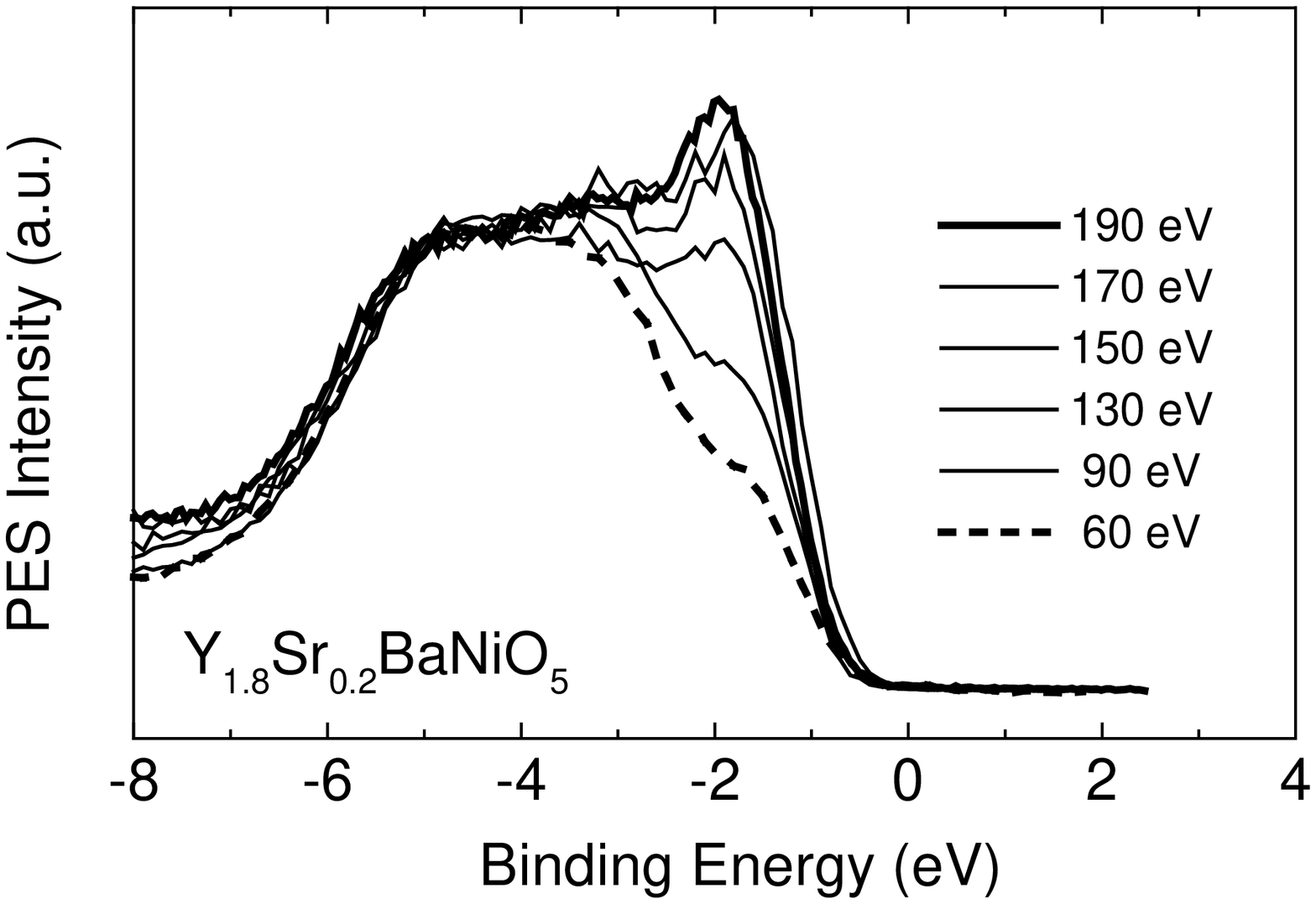}}
\end{center}
\caption{\label{two}Angle-integrated photoemission spectra of
$Y_{1.8}Sr_{0.2}BaNiO_5$ for several photon energies in the 60
eV-200 eV range.}
\end{figure}

\begin{figure}
\begin{center}
\scalebox{0.4}{\includegraphics{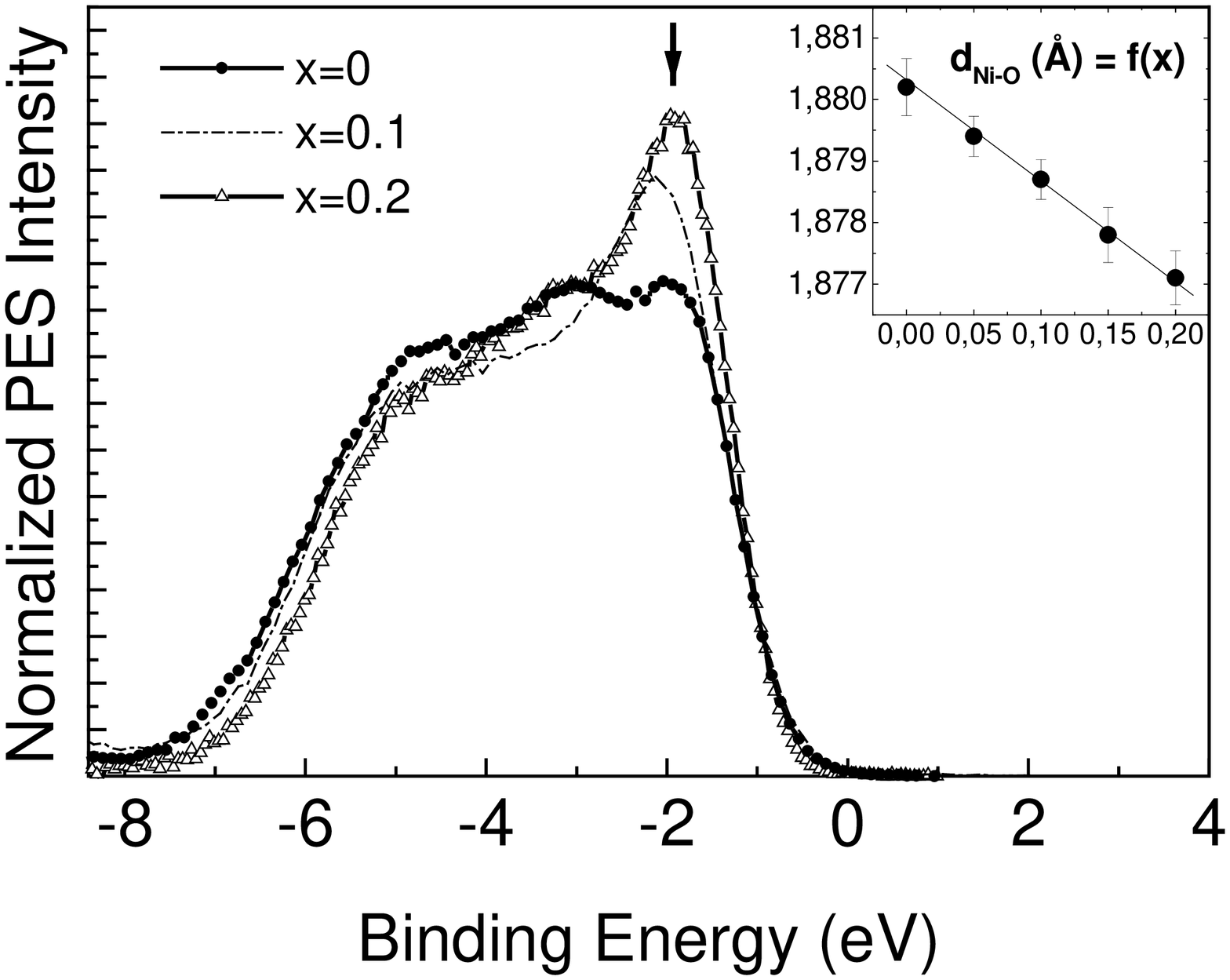}}
\end{center}
\caption{\label{three}Angle-integrated Photoemission spectra for
several doping after substraction of the Shirley background and
normalized to the area (see explanation in the text). Insert: Ni-O
distance as a function of doping x for Sr substituted samples.
Data taken from Ref.\cite{jalcom}}
\end{figure}

\begin{figure}
\begin{center}
\scalebox{0.5}{\includegraphics{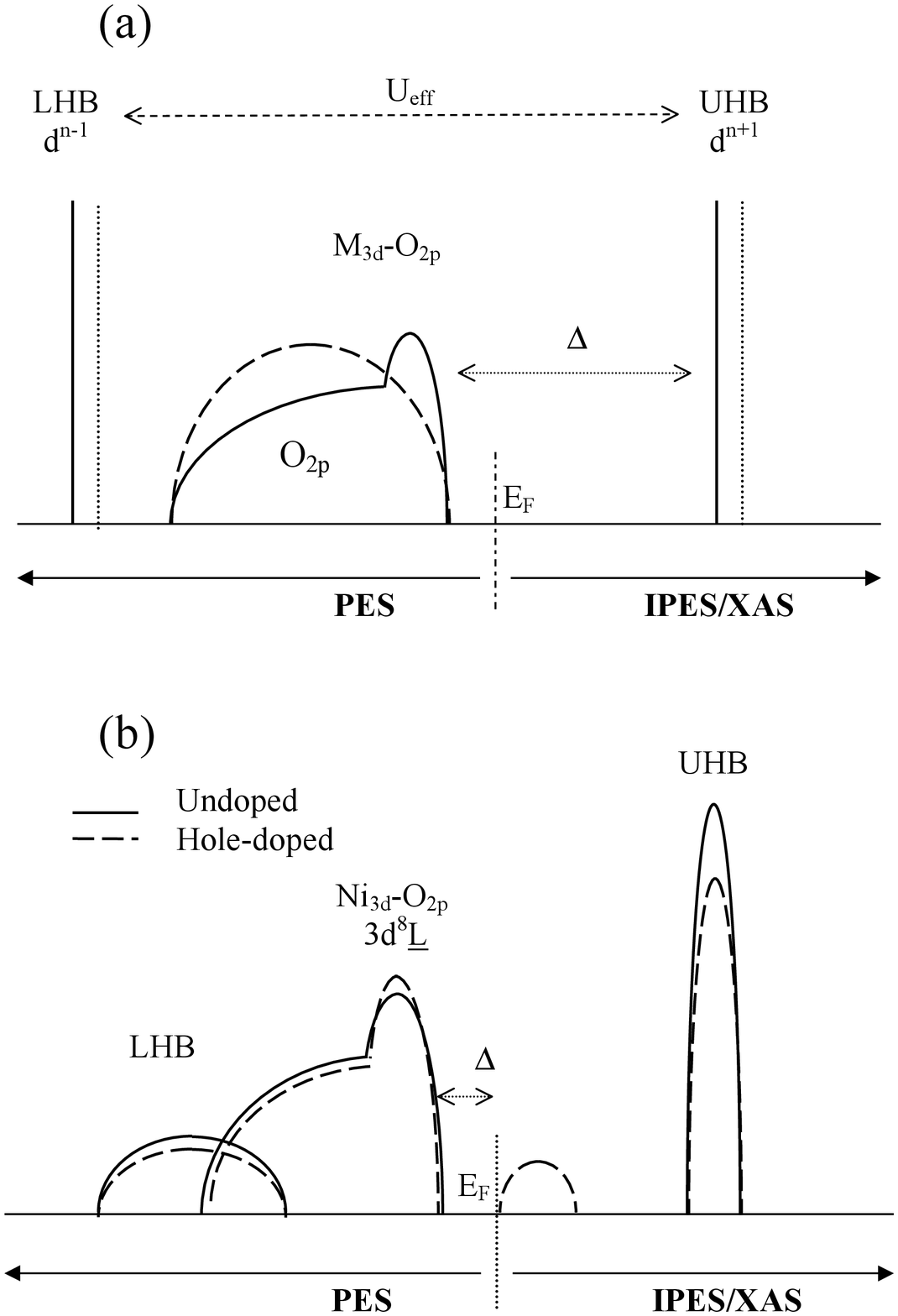}}
\end{center}
\caption{\label{four}(a) Effect of hybridization on the spectral
density of a non-doped charge-transfer oxide in the impurity
approach. From Ref.\cite{saw}. (b)Schematic view of the situation
encountered in $Y_{2-x}Sr_xBaNiO_5$. Hole doping leads to a large
spectral weight transfer in the electron-addition part and more
surprisingly to a spectral weight increase in the first
electron-removal states.}
\end{figure}

Alternatively, transition metal oxides can be described in many
body approaches\cite{saw}. $Y_2BaNiO_5$ belongs to the charge
transfer regime in the ZSA scheme because the charge transfer
parameter ($\Delta$) is smaller than the Coulomb term U
\cite{mai2}. In this regime, the ligand p band is situated between
the two Hubbard subbands. However, the photon energy dependence of
the photoemission spectrum clearly shows that Ni-3d symmetry
dominates in the first excited states of the photoemission
spectrum in contrast to naively expected in the charge transfer
regime for which the first electron removal excitations correspond
to formation of ligand holes in the p band. This behavior results
from hybridization effect. Indeed, as shown in the single impurity
model\cite{saw}, the hybridization between Ni-3d and O-2p leads to
a modification of the simple picture of a O-2p band between the
lower and upper Hubbard bands. The filled lower band will
hybridize with the O-2p states inducing the appearance of an image
Hubbard band at the low energy side of the p
band\cite{saw,esk}(figure \ref{four}(a)). Cluster calculation
shows that the first excitation state has a dominant
$d^8\underline{L}$ character\cite{mai2} confirming that the
photoemission hole is essentially on the ligand site. Such a
behavior is usually encountered in divalent Ni oxide like
NiO\cite{fuji}. Figure \ref{two} shows the same photon energy
dependence for the doped $Y_{1.8}Sr_{0.2}BaNiO_5$ compound. This
evolution shows that the spectral feature at low energy
significantly increases in intensity with increasing photon energy
confirming that this feature has a Ni-3d character. Comparison
with figure \ref{one} shows that the main effect of hole doping is
to increase the intensity of feature A. Moreover, a narrowing of
the bandwidth is also observed. In order to illustrate the
increase in intensity, we report in figure \ref{three} the
photoemission spectra of x=0.0, 0.1 and 0.2 oxides for $h\nu$ =190
eV after substraction of the standard Shirley background. Note
that this photon energy has been chosen in order to emphasize the
doping effect on the spectral density. Nevertheless, figure 1 and
figure 2 show that this effect is not energy dependent. The figure
3 illustrates the large effect of hole doping on the photoemission
spectral weight in $Y_{2-x}Sr_xBaNiO_5$. First, the doping
induced-states have negligible density of states near the Fermi
energy and do not yield the formation of a metal. The
photoemission-probed part of the gap (400 meV) seems indeed to be
not affected by doping. The present data and the optical
conductivity of reference \cite{ito}, which evidences a gap of 0.3
eV, confirm that the Fermi level is located immediately below the
new unoccupied states induced by doping.

The absence of spectral weight at $E_F$ is corroborated by
resistivity measurements which show that the doped oxides remain
insulating\cite{ito}. Secondly, even for low x, the A feature
exhibits a very large increase in intensity with doping. This
behavior contrasts to the usually observed spectral weight
modification in hole doped divalent Ni oxides, and more generally
in cuprates and nickelates. In $La_{2-x}Sr_xNiO_4$, for instance,
no significant increase in the spectral weight intensity is
observed for doping level smaller than x=0.5 \cite{eis}.
Unfortunately, due to the relative content of O-2p and Ni-3d
states, it is not really possible to quantitatively estimate the
weight of the $d^8\underline{L}$ state. Transfers of spectral
weight in hole-doped strongly correlated systems have been
previously discussed in detail in the literature\cite{esk}.
Anomalous spectral weight modifications related to electronic
correlations are expected in the electron-addition part
(unoccupied states) of hybridized charge transfer systems. This
trend is indeed observed in the XAS oxygen K edge of
$Y_{2-x}(Ca,Sr)_xBaNiO_5$ \cite{dit,jalcom} as well as in inverse
photoemission\cite{mai2}. However the electron-removal part of the
spectrum (occupied states) should slightly decrease upon doping
within this approach (since some low energy states are
depopulated) in contrast to the spectroscopic behavior.
An additional mechanism has to be invoked to explain the evolution
of photoemission spectra with Sr content. Interestingly, hole
doping in $Y_2BaNiO_5$ could result in an increase in the
$Ni3d-O2p$ hybridization since, unlike in
$La_{2-x}Sr_xNiO_4$\cite{Heaney}, the average Ni-O distance
shortens continuously upon doping (see insert figure 3 for
$Y_{2-x}Sr_xBaNiO_5$ and references \cite{Alonso,Massaroti} for
$Y_{2-x}Ca_xBaNiO_5$). As the apical Ni-O distance in $Y_2BaNiO_5$
is the shortest one observed in nickelates, a small decrease
should lead to a large increase in hybridization. This will be
emphasized in case of an inhomogeneous doping. If the doped holes
are trapped in local bound states, disturbing the electronic
states only in their vicinity \cite{xu,ito}, the real Ni-O bonds
in these perturbated part of the sample should be shorter than the
average one measured by X-rays. As a consequence, in the
photoemission spectra, which are the simple addition of
unperturbated parts and doped parts, the A peak enhancement should
originate mainly from the hole-doped regions where the
hybridization strongly increases. Therefore, the d character of
the top of the valence band should be emphasized in the doped
compounds due to this strong hybridization enhancement. A similar
effect is expected for Ca doped materials and has been observed
for highly doped samples in X-ray photoemission \cite{mai1} in
spite of the poor energy resolution.

This scenario of a very large spectral weight modification due to
doping-dependent hybridization is in qualitative agreement with
the modification of the valence photoemission spectra we observed
with Sr content in $Y_{2-x}Sr_xBaNiO_5$. Figure \ref{four}
summarizes the spectroscopic results obtained on hole doped
$Y_{2-x}Sr_xBaNiO_5$. Firstly, we recall in Fig. \ref{four}(a) the
effect of hybridization in the impurity approach used to describe
qualitatively the spectroscopy of homogeneous charge transfer
transition metal oxides\cite{saw}. Hybridized $d^8\underline{L}$
states appear at the top of the O-2p band. Obviously, the 3d
character of these states increases with increasing Ni3d-O2p
hybridization. Secondly, Fig.\ref{four}(b) illustrates the effect
of hole doping in $Y_{2}BaNiO_5$. In charge transfer systems, hole
doping usually affects the electron-addition spectra by the
appearance of the spectral feature just above the Fermi energy
associated with the hole states. Hybridization, by introducing d
character in these states, could enhance its spectral weight.

The singular behavior in doped $Y_{2}BaNiO_5$ is the increase in
intensity of the first ionisation states. This evolution, similar
to what is presented in Fig. \ref{four}(a), reflect the increase
in hybridization within the localized hole-doped regions. We note
that a theoretical treatment, based on a multiband Hamiltonian
containing the relevant Ni and O orbitals, gives large values of
transfer integrals within a hole-doped $NiO_6$ cluster
\cite{Batista98}. This singular spectroscopic signature, only
observed in doped $Y_2BaNiO_5$, could result from the particularly
small Ni-O distance characterizing this quasi-1D compound.

%
%
\section{conclusion}

To summarize, we have performed an angle-integrated photoemission
investigation of the spectral weight modification by hole doping
in the $Y_{2-x}Sr_xBaNiO_5$ oxides. In this Haldane chain
compound, hole doping leads to strong modifications in the
unoccupied part of the spectral density as usually observed in
cuprates and nickelates. However, photoemission spectroscopy
exhibits a singular behavior since a significant spectral weight
increase corresponding to final states with d character is
observed. We propose that this surprising behavior could result
from an increase in hybridization between Ni-3d and O-2p states
induced by the reduction of Ni-O distance in the chain direction
by doping. Such an effect is favoured by the inhomogeneous
character of doping. We hope that reliable calculations will be
performed in the future to understand this anomalous spectral
weight enhancement.
\section{Acknowledgement}
We would like to thank M. Grioni for helpful discussion and A.
Taleb for her hospitality on the SU3 beamline.
%
%


\begin{thebibliography}{99}

\bibitem{Amador}J. Amador, E. Gutierrez-Puebla, M.A. Monge,
I. Rasines, C. Ruiz-Valero, F. Fernandez, R. Saez-Puche and J.A.
Campa, \prb {\bf 42}, 7918 (1990).
\bibitem{mattheiss}L.F. Mattheiss, \prb {\bf 48}, 4352 (1993).
\bibitem{dar}J. Darriet and L.P. Regnault, Solid State Commun. {\bf
856}, 409 (1993).
\bibitem{affl}I. Affleck, T. Kennedy, E.H. Lieb and H.
Tasaki, \prl {\bf 59}, 799 (1987).
\bibitem{dit}J.F. Ditusa, S.W. Cheong, J.H. Park, G. Aeppli, C. Broholm,
and C.T. Chen, \prl {\bf 73}, 1857 (1994).
\bibitem{dag}E. Dagotto, J. Riera, A. Sandvik and A. Moreo,
\prb {\bf 76}, 1731 (1996).
\bibitem{xu}JG. Xu, G. Aeppli, M.E. Bisher, C. Broholm, J.F.
Ditusa,C.D. Frost, T. Ito, K. Oka, R.L. Paul, H. Takagi and M.M.J.
Treacy, Science {\bf 289} (2000).
\bibitem{zaa}J. Zaanen, G.A. Sawatzky, and J.W. Allen, \prl {\bf 55}, 418 (1985).
\bibitem{ito}I. Ito, H. Yamaguchi, K. Oka, K.M. Kojima, H. Eisaki and
S. Uchida, \prb {\bf 64}, 060401(R) (2001).
\bibitem{mai2}K. Maiti, P. Mahadevan and D.D. Sarma, \prb {\bf 59}, 12457 (1999).
\bibitem{jalcom}F.-X. Lannuzel, E. Janod, C. Payen, G. Ouvrard, P. Moreau, O. Chauvet, P. Parent,
C. Laffon, J. Alloys Comp. {\bf 317-318}, 149 (2001).
\bibitem{mai1}K. Maiti and D.D. Sarma, \prb {\bf 58}, 9746 (1998).
\bibitem{nov}P. Nov$\acute{a}$c F. Boucher, P. Gressier, P. Blaha, K. Schwarz \prb {\bf
63}, 235114 (2001).
\bibitem{fuji}A. Fujimori and F. Minami,  \prb {\bf 30},
957 (1984).
\bibitem{saw}J. Zaanen and G.A. Sawatzky, J. Solid State Chem. 88, {\bf 8}, 8 (1990);
Prog. Theor. Phys. {\bf Suppl. 101} 231(1990).
\bibitem{esk}H.Eskes, M.B. J. Meinders and G.A. Sawatzky, \prl {\bf 67}, 1035 (1991).
\bibitem{eis}H. Eisaki, S. Uchida, T. Mizokawa, H. Namatame, A. Fujimori, J. van Elp,
P. Kuiper, G.A. Sawatzky, S. Hosoya and H. Katayama-Yoshida, \prb
{\bf 45}, 12513 (1992).
\bibitem{Heaney}P.J. Heaney, A. Mehta, G. Sarosi, V.E. Lamberti,
A. Navrotsky, \prb {\bf 57}, 10370 (1998). In this work, a careful
analysis of the position of the oxygen in the plane of
$La_{2-x}Sr_xNiO_4$ indicates that despite the shortening of the
cell parameters, the Ni-O distance may even increase due to a
rotation of the $NiO_6$ octahedra.
\bibitem{Alonso}J.A. Alonso, I. Rasines, J. Rodriguez-Carvajal and J.B. Torrance,
 J. Solid State Chem., {\bf 109}, 231 (1994).
\bibitem{Massaroti}V. Massaroti, D. Capsoni, M. Bini, A. Altomare and A.G. Moliterni,
Z. Kristallogr. {\bf 214}, 231 (1999).
\bibitem{Batista98} C.D. Batista, A.A. Aliglia and J. Eroles,
Europhys. Lett. {\bf 43}, 71 (1998).
\end{thebibliography}
\end{document}